# The Importance of Being Negative:

# A serious treatment of non-trivial edges in brain functional connectome

Liang Zhan[1*], Lisanne Michelle Jenkins[2], Ouri Wolfson[3], Johnson Jonaris GadElkarim[4], Kevin Nocito[4], Paul M. Thompson[5], Olusola Ajilore[2], Moo K Chung[6], Alex Leow[2, 4*]

[1]Computer Engineering Program, University of Wisconsin-Stout, Menomonie, WI, USA, 54751

[2]Department of Psychiatry, University of Illinois, Chicago, IL, USA, 60607

[3]Department of Computer Science, University of Illinois, Chicago, IL, USA, 60607

[4]Department of Bioengineering, University of Illinois, Chicago, IL, USA, 60607

[5]Imaging Genetics Center, and Institute for Neuroimaging and Informatics, Keck School of Medicine of USC, CA, USA, 90032

[6]Department of Biostatistics and Medical Informatics, University of Wisconsin-Madison, Madison, WI, USA, 53706

* Corresponding author:

Alex Leow:  aleow@psych.uic.edu.   Liang Zhan: zhanl@uwstout.edu

## Abstract

Understanding the modularity of fMRI-derived brain networks or 'connectomes' can inform the study of brain function organization. However, fMRI connectomes additionally involve negative edges, **which may not be optimally accounted for by existing approaches to modularity that variably threshold, binarize, or arbitrarily weight these connections. Consequently, many existing Q maximization-based modularity algorithms yield variable modular structures.** Here we present an alternative **complementary** approach that exploits how frequent the BOLD-signal correlation between two nodes is negative. We validated this novel probability-based modularity approach on two independent publicly-available resting-state connectome datasets (the Human Connectome Project and the 1000 Functional Connectomes) and demonstrated that negative correlations alone are sufficient in understanding resting-state modularity. In fact, this approach a) permits a dual formulation, leading to equivalent solutions regardless of whether one considers positive or negative edges; b) is theoretically linked to the Ising model defined on the connectome, thus yielding modularity result that maximizes data likelihood. Additionally, we were able to detect novel and consistent sex differences in modularity in both datasets. As datasets like HCP become widely available for analysis by the neuroscience community at large, **alternative and perhaps more advantageous** computational tools to understand the neurobiological information of negative edges in fMRI connectomes are increasingly important.

**Keywords:** Brain, fMRI, Resting-State, Negative BOLD-signal Correlation, Connectome, Modularity, Community, Dual-form, Network analysis

# 1. Introduction

Just as social networks can be divided into cliques that describe modes of association (e.g. family, school), the brain's connectome can be divided into modules or communities. Modules contain a series of nodes that are densely interconnected (via edges) with one another but weakly connected with nodes in other modules (Meunier et al., 2010). Thus, modularity or community structure best describes the intermediate scale of network organization, rather than the global or local scale. In many networks, modules can be divided into smaller sub-modules, thus can be said to demonstrate hierarchical modularity and *near decomposability* (the autonomy of modules from one another), a term first coined by Simon in 1962 (Simon, 2002, Meunier et al., 2010). Modules in fMRI-derived networks comprise anatomically and/or functionally related regions, and the presence of modularity in a network has several advantages, including greater adaptability and robustness of the function of the network. Understanding modularity of brain networks can inform the study of organization and mechanisms of brain function and dysfunction, thus potentially the treatment of neuropsychiatric diseases.

Mathematical techniques derived from graph theory (Fornito et al., 2013) have been developed to measure and describe the modular organization of neural connectomes (Bullmore and Sporns, 2009, Sporns and Betzel, 2016). Different methods for module detection have been applied in network neuroscience, and offer different strengths and weaknesses (reviewed in Sporns and Betzel, 2016). Optimization algorithms are typically used to maximize the *Q* modularity metric or its variants (Danon et al., 2005). These algorithms vary in accuracy as there are tradeoffs made with computational speed (Rubinov and Sporns, 2010). Simulated annealing (e.g. Guimera et al., 2004, Guimera and Amaral, 2005) is a slower, more accurate method for smaller networks, however could be computationally expensive with larger networks (Danon et al., 2005). The Newman method (Newman and Girvan, 2004, 2006) reformulates modularity with consideration of the spectral properties of the network, and is also considered fairly accurate with adequate speed for smaller networks (Rubinov and Sporns, 2010). More recently, the Louvain method (Blondel et al., 2008) has been developed for large networks (millions of nodes and billions of edges). Its rapid computation and ability to detect modular hierarchy (Rubinov and Sporns, 2010) has led to it becoming one of the most widely utilized methods for detecting communities in large networks. Comparisons with other modularity optimization methods have found that the Louvain method outperforms numerous other similar methods (Lancichinetti and Fortunato, 2009, Aynaud et al., 2013).

**However, these existing methods were mostly originally developed for networks with only positive connections and may additionally suffer from suboptimal reproducibility** (Butts, 2003, Fortunato and Barthelemy, 2007, Guimera and Sales-Pardo, 2009). With the advent of connectomics, they also have been heuristically applied to fMRI brain networks, in which we have the additional complication of negative correlations. To this end, **some** methods largely ignore fMRI networks' negative edges (Fornito et al., 2013), only considering the right tail of the correlation histogram, i.e. the positive edges (Schwarz and McGonigle, 2011). However, in functional neuroimaging, negative edges may be

neurobiologically relevant (Sporns and Betzel, 2016), depending on factors such as data preprocessing steps, particularly the removal of potentially confounding signal such as head motion, global white-matter or whole-brain average signal, before calculation of the correlation matrix, because removal of such signal could result in detection of anticorrelations that were not present in the original data (Schwarz and McGonigle, 2011). Ignoring negative edges is achieved with binarization of a network (so-called 'hard thresholding'), by selecting a threshold then replacing edge values below this threshold with zeros, and replacing supra-threshold values with ones (van den Heuvel et al., 2017). Some researchers retain the weights of the supra-threshold edge values, which has the effect of compressing the positive edges, however the negative edges remain suppressed (Schwarz and McGonigle, 2011). Choice of threshold is important as more severe thresholds increase the contributions from the strongest edges, but can result in excessive disconnection of nodes within networks, in comparison to less stringent thresholds. Rather than binarizing networks, some researchers choose a 'soft thresholding' approach that replaces thresholding with a continuous mapping of correlation values into edge weights, which had the effect of suppressing, rather than removing weaker connections (Schwarz and McGonigle, 2011). Linear and non-linear adjacency functions can be employed, and the choice can be made to retain the valence of the edge weights, when appropriate.

An alternative to optimization methods discussed above, Independent Components Analysis (ICA) has been applied to functional neuroimaging data (Beckmann et al., 2005). This method assumes that voxel time series are linear combinations of subsets of representative time series (Sporns and Betzel, 2016). Patterns of voxels load onto spatially independent components (modules). Unlike optimization methods, ICA allows for overlapping communities (Sporns and Betzel, 2016), although the number of ICA components needs to be pre-specified.

Utilizing a distance-based approach, recently, a new technique for investigating the hierarchical modularity of structural brain networks has been developed (GadElkarim et al., 2012, GadElkarim et al., 2014). Rather than maximizing $Q$, the Path Length Associated Community Estimation (PLACE) uses a unique metric **that** measures the difference in path length between versus within modules, to both maximize within-module integration and between-module separation (GadElkarim et al., 2014). It utilizes a hierarchically iterative procedure to compute global-to-local bifurcating trees (i.e. dendrograms), each of which represents a collection of nodes that form a module.

In this study, we **developed a related novel** method for functional brain networks – Probability Associated Community Estimation (PACE), that uses probability, not thresholds or the magnitude of BOLD signal correlations. **We conducted experiments using this method, as well as six different implementations within the widely used Brain Connectivity Toolbox (BCT) (http://www.brain-connectivity-toolbox.net/) using data** from the freely accessible 1000 Functional Connectomes or F1000 Project dataset (Biswal et al., 2010) and the Human Connectome Project (HCP) (Van Essen et al., 2012, Van Essen et al., 2013), **and further** examined differences in resting-state functional connectome's modularity (i.e., the resting-state networks or RSN) between males and females.

## 2. Methods

The popular *Q*-based modular structure (Reichardt and Bornholdt, 2006, Blondel et al., 2008, Ronhovde and Nussinov, 2009, Sun et al., 2009, Rubinov and Sporns, 2011) is extracted by finding the set of non-overlapping modules that maximizes the modularity metric *Q*:

$$Q(G) = \frac{1}{2m} \sum_{i \neq j} \left( A_{ij} - \frac{k_i k_j}{2m} \right) \delta(i,j)$$

For a binary graph G, m is the total number of edges, $A_{ij}$ = 1 if an edge links nodes *i* and *j* and 0 otherwise, $\delta(i,j)$ = 1 if nodes *i* and *j* are in the same community and 0 otherwise, and $k_i$ is the node degree of *i* (i.e., its number of edges). For weighted graphs, m is the sum of the weights of all edges while $A_{ij}$ becomes the weight of the edge that links nodes *i* and *j* and $k_i$ the sum of all weights for node *i*.

**Approaches based on Q-maximization** are naturally suitable for understanding the modularity of structural connectome where all edges are non-negative. As an alternative to Q maximization, we previously developed a graph distance (shortest path length) based modularity approach for the structural connectome. By exploiting the structural connectome's *hierarchical modularity*, this *path length associated community estimation technique* (PLACE) is designed to extract global-to-local hierarchical modular structure in the form of bifurcating dendrograms (GadElkarim et al., 2012). PLACE has potential advantages over *Q* (GadElkarim et al., 2014), as it is hierarchically regular and *scalable* by design.

Here, the degree to which nodes are separated is measured using graph distances (Dijkstra, 1959) and the PLACE benefit function is the $\Psi^{PL}$ metric, defined at each bifurcation as the difference between the mean inter- and intra- modular graph distances. Thus, maximizing $\Psi^{PL}$ is equivalent to searching for a partition with stronger intra-community integration and stronger between-community separation (GadElkarim et al., 2012, GadElkarim et al., 2014, Ye et al., 2015, Lamar et al., 2016, Zhang et al., 2016).

**2.1 Probability-associated community estimation (PACE) for functional connectomes.**

Here let us describe the PACE-based modularity of a functional connectome mathematically represented as an undirected graph $FC(V,E)$, where *V* is a set of vertices (i.e., nodes) and *E* is a set of edges (indexed by considering all pairs of vertices). Each edge of *E* is associated with a weight that can be either positive or negative.

Given a collection of functional connectomes *S* on the same set of nodes *V* (but having edges with different weights), we can define the following *aggregation graph G (V, E)*. For each edge $e_{i,j}$ in *E* connecting node *i* and node *j*, we consider $P^{-}_{i,j}$, the probability of observing a negative value at this edge in *S* (i.e., the BOLD signals of *i* and *j* are anti-correlated). In the case of HCP, for example, *S* thus consists of all healthy subjects' resting-state functional connectome and this probability can simply be estimated using the ratio between the number of connectomes in *S* having the edge $e_{i,j} < 0$ and the total number of connectomes in *S*. Similarly, we define the probability of an edge in *E* being non-negative as $P^+_{i,j}$. Naturally, the $P^-$- $P^+$ pair satisfies the following relationship:

$$P^-_{i,j} + P^+_{i,j} = 1, \quad \forall (i,j), i \neq j$$

Then, given $C^1$, $C^2$,…, $C^N$ that are *N* subsets (or communities) of *V*, we define the mean intra-community edge positivity or negativity $\overline{P^\pm(C^n)}$ for the *n*-th community $C^n$ as:

$$\overline{P^\pm(C^n)} = \frac{\sum_{i,j \in C^n, \ i<j} P^\pm_{i,j}}{|C^n|(|C^n|-1)/2}$$

Here $|C^n|$ represents the size (i.e., number of nodes) of the *n*-th community. Similarly, we could define the mean inter-community edge positivity and negativity (between communities $C^n$ and $C^m$) as:

$$\overline{P^\pm(C^n, C^m)} = \overline{P^\pm(C^m, C^n)} = \frac{\sum_{i \in C^n, j \in C^m} P^\pm_{i,j}}{|C^n||C^m|}$$

Here, the first equality holds as correlation-based functional connectomes are undirected. The intuition of PACE for fMRI connectomes is that edges that are most frequently anti-correlations should be placed across communities.

PACE operates as follows. Given a collection of functional connectomes *S*, PACE identifies a natural number *N* and a partition of *V*, $C^1 \cup C^2 \cup ... \cup C^N = V, (C^i \cap C^j = \emptyset \ for \ all \ i \neq j$) which maximizes the PACE benefit function Ψ. Intuitively, Ψ computes the difference between mean inter-community edge negativity and mean intra-community edge negativity. Moreover, considering the duality between $P^-$ and $P^+$, our optimization problem thus permits an equivalent dual form. Formally, Ψ=

$$\underset{C^1 \cup C^2 \cup \ldots \cup C^N = V, C^i \cap C^j = \emptyset \text{ for all } i \neq j}{\text{argmax}} \left\{ \frac{\sum_{1 \leq n < m \leq N} \overline{P^-(C^n, C^m)}}{N(N-1)/2} - \frac{\sum_{1 \leq n \leq N} \overline{P^-(C^n)}}{N} \right\} =$$

$$\underset{C^1 \cup C^2 \cup \ldots \cup C^N = V, C^i \cap C^j = \emptyset \text{ for all } i \neq j}{\text{argmax}} \left\{ \frac{\sum_{1 \leq n \leq N} \overline{P^+(C^n)}}{N} - \frac{\sum_{1 \leq n < m \leq N} \overline{P^+(C^n, C^m)}}{N(N-1)/2} \right\}$$

To solve the above PACE optimization problem, we adopt a PLACE-like algorithm, which has been extensively validated (GadElkarim et al., 2012, Ajilore et al., 2013, GadElkarim et al., 2014, Ye et al., 2015, Lamar et al., 2016, Zhang et al., 2016), and computed global-to-local 4-level bifurcating trees (yielding a total of 16 communities at the fourth level; please refer to GadElkarim et al. (2014) for implementation details).

2.2 Theoretical link between PACE and the Ising model

Here let us further explore the relationship between PACE and the Ising model using a mean-field approximation approach. When defined on the human connectome, the Ising model consists of assigning atomic spins $\sigma$ to each brain region or node to one of two states (+1 or −1). Given a specific ensemble spin configuration $\sigma$ over the entire brain and assuming the absence of external magnetic field, the corresponding Hamiltonian is thus defined as:

$$H(\sigma) = -\sum_{(i,j) \in E} J_{ij} \sigma_i \sigma_j$$

Here $(i,j) \in E$ indicates that there is an edge connecting nodes $i$ and $j$. In classic thermodynamics, the Hamiltonian relates a configuration to its probability via the following Boltzmann distribution equation:

$$P(\sigma) = \frac{e^{-\beta H(\sigma)}}{Z}$$

Where $\beta$ is the inverse temperature and the normalizing constant Z is often called the partition function $Z = \sum_\sigma e^{-\beta H(\sigma)}$. Note, $J_{ij}$ is positive when the interaction is ferromagnetic $J_{ij} > 0$, and antiferromagnetic when $J_{ij} < 0$.

Given this general set-up, we are now ready to show the general equivalence between PACE and maximizing the joint likelihood of the observed rs-fMRI connectome data over some unknown ferromagnetic/antiferromagnetic ensemble interaction *J* defined on the connectome.

First, as one subject's rs-fMRI connectome is independent of other subjects', the joint likelihood over *S* subjects can be computed by forming the product:

$$Likelihood\ (oberseved\ rs-fMRI\ data\ |\ J) = \frac{e^{-\beta \sum_{s=1}^{S} H(\rho^s)}}{Z} = \frac{e^{-\beta S \sum_{s=1}^{S} \overline{H(\rho)}}}{Z}$$

For convenience of notations, let us work with the negative mean Hamiltonian

$$-\overline{H(\rho)} = -\frac{\sum_{s=1}^{S} H(\rho^s)}{S} = \sum_{(i,j)\in E} J_{ij} \frac{\sum_{s=1}^{S} \sigma_i^s \sigma_j^s}{S} = \sum_{(i,j)\in E} J_{ij} \overline{\sigma_i \sigma_j}$$

While the spin $\sigma_i^s$ at a node *i* for any subject *s* is unknown, with PACE we nevertheless could proceed to estimate the expected value of spin product (across all *S* subjects) $\overline{\sigma_i \sigma_j}$ by noting that PACE assigns $\sigma_i \sigma_j$=1 with probability $P^+{}_{i,j}$ and -1 with probability $P^-{}_{i,j}$. Thus

$$\overline{\sigma_i \sigma_j} = 1 \cdot P^+{}_{i,j} + (-1) \cdot P^-{}_{i,j} = 1 \cdot P^+{}_{i,j} + (-1) \cdot (1 - P^+{}_{i,j}) = 2P^+{}_{i,j} - 1$$

$$= 1 - 2P^-{}_{i,j}$$

Next, let us compute and simplify $\overline{H(\rho)}$ using the above equations, coupled with mean-field approximation, by separately considering ferromagnetic vs. antiferromagnetic interactions (i.e., with respect to the sign of $J_{ij}$):

$$J^+ = \frac{\sum_{(i,j)\in E, J_{ij}>0} J_{ij}}{|(i,j)\in E, J_{ij}>0|} > 0; \quad J^- = -\frac{\sum_{(i,j)\in E, J_{ij}<0} J_{ij}}{|(i,j)\in E, J_{ij}<0|} > 0$$

Thus

$$-\overline{H(\rho)} = \sum_{(i,j)\in E} J_{ij}\overline{\sigma_i\sigma_j} = \sum_{(i,j)\in E, J_{ij}>0} J_{ij}\overline{\sigma_i\sigma_j} + \sum_{(i,j)\in E, J_{ij}<0} J_{ij}\overline{\sigma_i\sigma_j}$$

$$\approx J^+ \sum_{(i,j)\in E, J_{ij}>0} \overline{\sigma_i\sigma_j} - J^- \sum_{(i,j)\in E, J_{ij}<0} \overline{\sigma_i\sigma_j}$$

$$= J^+ \sum_{(i,j)\in E, J_{ij}>0} (2P^+{}_{i,j} - 1) + J^- \sum_{(i,j)\in E, J_{ij}<0} (2P^-{}_{i,j} - 1)$$

Note, since here "mean-field" is constructed by averaging over ferromagnetic/antiferromagnetic interaction terms, our formulation may instead be called a mean-interaction approach.

Last, realizing that maximizing the joint likelihood of the observed data with respect to unknown ensemble interactions *J* (and thus unknown mean-interaction approximations $J^+$ and $J^-$) is equivalent to maximizing the negative mean Hamiltonian $-\overline{H(\rho)}$, we examine the right-hand side of the above equation (and note that both $J^+$ and $J^-$ are non-negative) and deduct that a general maximization strategy can be devised by:

*a) assigning as much as possible any two nodes i, j that are highly likely to exhibit positive BOLD correlation $e_{i,j} > 0$ (and thus the term $(2P^+{}_{i,j} - 1)$ more likely to be positive) to have ferromagnetic interactions (thus i, j more likely to be placed in the same community),* and at the same time

*b) assigning as much as possible nodes i, j that are highly likely to exhibit negative BOLD correlation $e_{i,j} < 0$ (and thus the term $(2P^-_{i,j} - 1)$ more likely to be positive) to have antiferromagnetic interactions (thus i and j more likely to be placed in different communities).*

This is exactly the intuition of PACE, i.e., we maximize *the difference between mean inter-community edge negativity and mean intra-community edge negativity (or equivalently maximizing the difference between mean intra-community edge positivity and mean inter-community edge positivity).*

**2.3 Relaxation of the powers-of-two constraint: constructing the PACE null model and testing the statistical significance of each bifurcation**

**As PACE attempts, for each branch at a specific PACE level, to further split nodes within that branch into 2 subsequent groups, it is thus natural to ask if a procedure can be constructed in order to determine the level of statistical significance for such a split. By stopping a branch from further splitting when there is evidence against it, PACE can in theory yield any number of communities (no longer restricted to powers of 2).**

**Here, we propose such a procedure by first constructing the null distribution based on the observed data. Indeed, we can sample the null distribution (i.e., there is no modular patterns of co-/anti- activation) of the PACE benefit function Ψ by first randomly permuting the pair $P^+_{i,j}$ / $P^-_{i,j}$ for all (i, j) (i.e., randomly exchanging edge positivity with negativity, or simply put a probability value is replaced by 1 minus this value), followed by re-**

**running PACE with shuffled edge positivity/negative. Then, at each split the actual Ψ achieved by the original data is compared to the Ψ values of the reshuffled data at the same PACE level; if the former lies within the top 5% of the latter, such a split is determined to be significant (P < 0.05).**

**In sum, using this data-informed permutation procedure we relax the powers-of-two constraint during PACE optimization, thus letting the observed data to inform us the statistically most meaningful number of modules. This number can now be any positive integer which is no longer constrained to be a power of two.**

**3. Results**

**3.1 Data description**

We tested our PACE framework on two publicly available connectome datasets (Biswal et al., 2010, Brown et al., 2012). The first one is a 986-subject resting state fMRI connectome dataset from the 1000 functional connectome project (17 subjects' connectomes were discarded due to corrupted files), downloaded from the USC multimodal connectivity database (http://umcd.humanconnectomeproject.org). The dimension of the network is 177x177. The 2$^{nd}$ dataset is 820 subjects' resting state fMRI connectome from the Human Connectome Project (released in December 2015, named as HCP900 Parcellation+Timeseries+Netmats,

https://db.humanconnectome.org/data/projects/HCP_900). The dimension of the network is 200x200, derived using ICA. For details of these two datasets, please refer to their respective websites and references.

**3.2 Simulation study**

Here, we created a 100x100 edge-negativity probability map, which contains five modules (each module has 20 nodes; Fig 1a). The edge negativity values within each module are uniformly randomly generated between 0 to 0.5 (less likely anti-activation within module) and the values across modules are uniformly randomly assigned from 0.5 to 1. Then, 3-level PACE was applied to generate 8 modules (Fig 1b), followed by sampling the null distribution of Ψ with 1000 permutations using the procedure described in Section 2.3.

Results indicated that PACE correctly recovered the 5-module ground truth, and the null distribution procedure indeed rejected any further splitting beyond 5 modules (Fig 1c; blue lines indicate statistically meaningful bifurcations). Fig 1d further shows the performance of PACE across different levels of noise (the exact procedure of how noise is applied is discussed in the supplemental material).

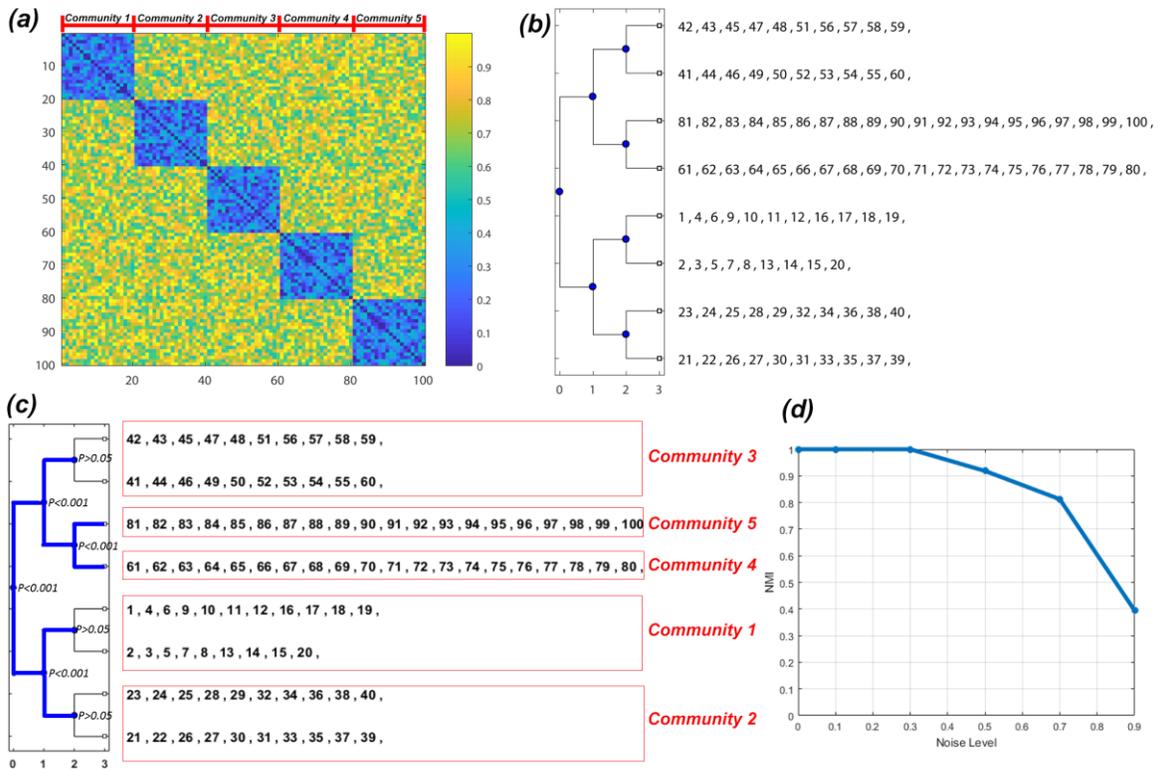

**Figure 1. Simulation study for PACE. (a) a 100 x 100 five-community edge-negative probability map was generated, where each module has 20 nodes. Within-community the edge negativity value is uniformly randomly selected from 0 to 0.5 and between-community the value from 0.5 to 1; (b) 3-level PACE was computed and 8 communities were generated; (c) using 1000 permutations randomly exchanging edge positivity with negativity we constructed the null distribution of Ψ, using which we tested the significance of each bifurcation. For the 4 possible bifurcations at the 3rd level PACE, only one was significant (significant bifurcations highlighted in blue), thus yielding a total of five modules, each of which matches the corresponding module in the ground truth (highlighted by the red square; from top to bottom, module correspondence is 3, 5, 4, 1, 2); (d) the performance of PACE in this toy example under different noise levels (L=0.1, 0.3, 0.5, 0.7 and 0.9; the exact procedure of how L is applied is discussed in the supplemental material).**

3.3 Stability Analysis

To better understand the stability of the PACE with respect to the number of subjects used in estimating edge negativity/positivity, we tested PACE on subsets of HCP and F1000 randomly generated with a bootstrapping procedure (Chung et al., 2006) (sampling with replacement) to investigate the reproducibility of the extracted community structure as a function of the sample size (from N=50 to 900 for F1000 and 50 to 800 for HCP; in increments of 50). For each N, 1000 bootstrap samples were generated and the resulting 1000 community structures were compared to the community structure derived from the entire HCP/F1000 sample using the normalized mutual information or NMI (Alexander-Bloch et al., 2012). NMI values are between 0 and 1, with 1 indicating two modular structures are identical. **Figure 2** illustrates the mean NMI (y-axis) as a function of sample size N (x-axis), for all levels of PACE. Careful examinations of these NMI values suggest that stable PACE-derived modularity can be obtained with as few as ~100 subjects **(note here we include all subjects, regardless of age and sex, during bootstrapping. it is likely that the NMI values would be even higher if we restrict the analysis to a narrower age range and/or one specific sex).**

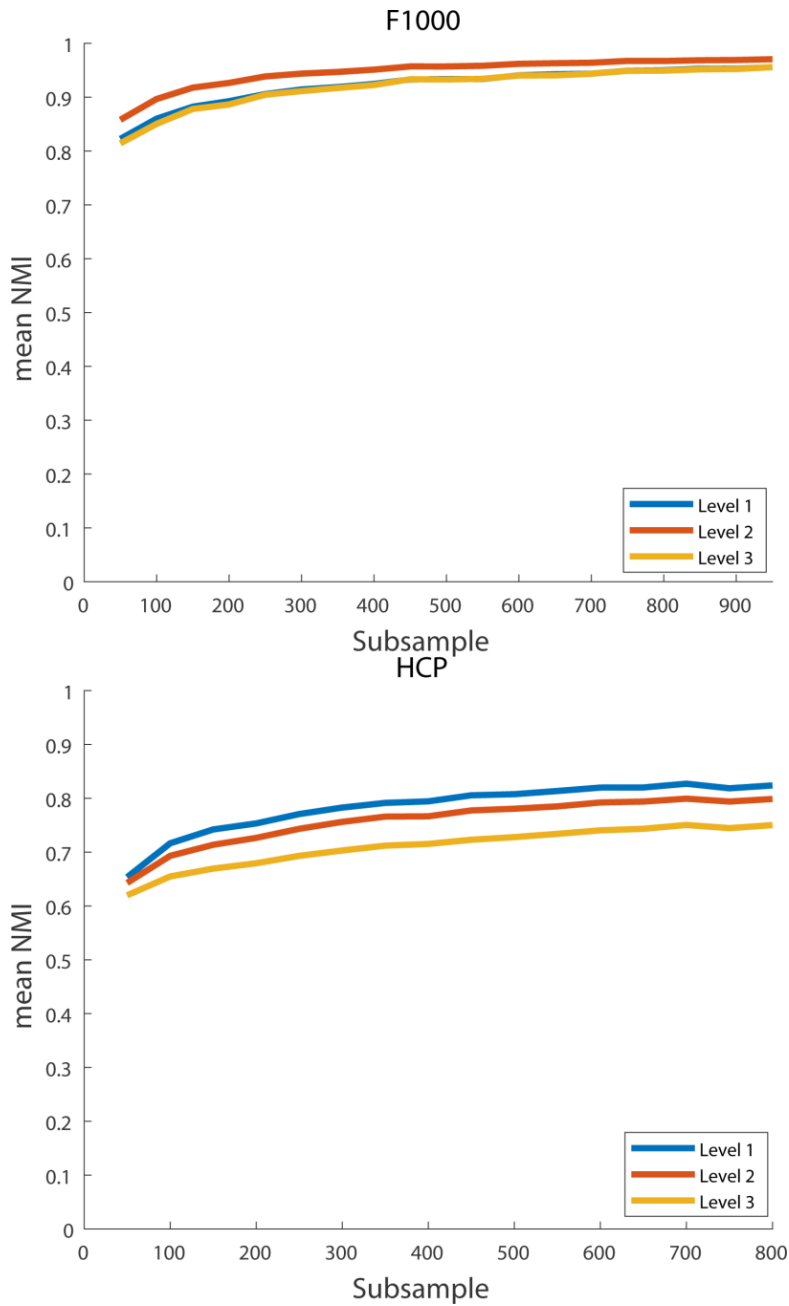

**Figure 2.** Stability of PACE as a function of the number of subjects used in estimating edge negativity (x-axis; N=50 to 900 for F1000 and N= 50 to 800 for HCP) using a bootstrapping procedure. For each N, we generated 1000 copies (random sampling with replacement) and computed the NMI between each of the 1000 corresponding PACE modular structures and that derived from the full sample. Y-axis plots the mean of these 1000 NMI values, for each N and each of PACE levels.

**3.4 Optimal number of PACE bifurcations informed by the null model in F1000 and HCP**

To determine the significance of PACE-derived hierarchical modularity at each bifurcation, for HCP and F1000 we generated 1000 samples of Ψ under the null hypothesis using the procedure in Section 2.3, and tested the significance of each split up to the fourth level. For both datasets, all bifurcations up to the 3$^{rd}$ level were significant. At the 4$^{th}$ level, in F1000 none of the 8 possible bifurcations was significant (thus resulting in a total 8 of modules) and in HCP only one of the 8 bifurcations was (p = 0.002, which remained significant after Bonferroni correction with a cut-off of 0.05/8), thus yielding a total of 9 modules. Fig. 3 illustrates the whole procedure in HCP. (Please refer to supplementary Fig. S2 for the final community structures for F1000 and HCP.)

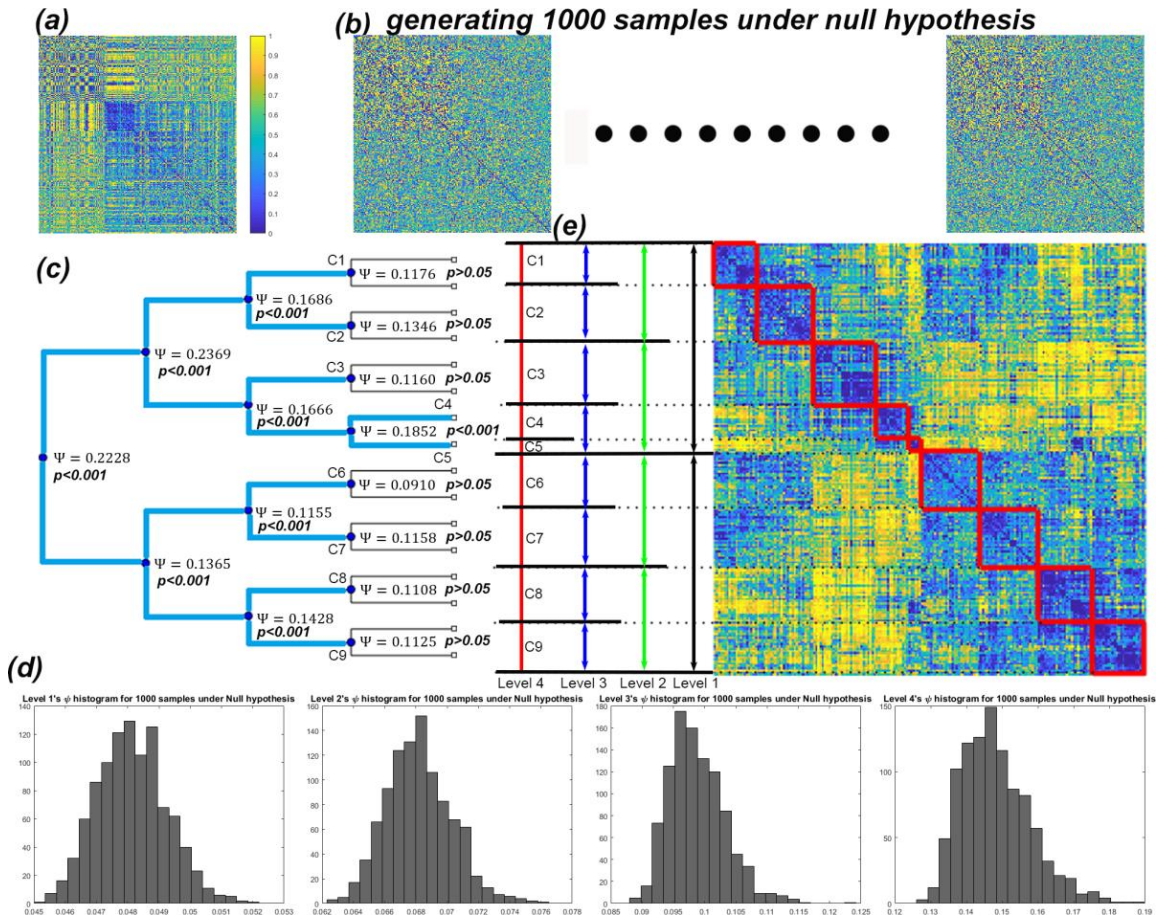

Figure 3. Constructing the PACE null model in HCP. (a) original HCP edge-negativity frequency map; (b) generating 1000 samples of the same map under the null hypothesis, by randomly permuting the edge positivity/negativity pair (i.e., for each element of the matrix, its edge-negativity value p is randomly re-assigned to 1-p with a probability of 50%); (c) Testing the significance of each PACE bifurcation in HCP up to the 4$^{th}$ level tree structure. The ψ values achieved by the original data are shown at each bifurcation point, along with their statistical significance. (d) Histograms of the 1000 PACE benefit function ψ values generated under the null model for each level. Only bifurcations with observed ψ values ranked in the top 50 (with respect to the 1000 null-model values from the same PACE level) is considered "significant" (blue lines in c). At the fourth level, only one out of eight possible bifurcations was meaningful, resulting in nine final communities (C1~C9). (e) re-arranged matrix to show how these communities (C1~C9) are formed from Level 1~Level 4.

## 3.5 Modular structures revealed using PACE versus weighted-Q maximization based methods

In this section, we compared PACE-derived modularity with Q-based modularity computed from the mean F1000 or HCP functional connectome (mean connectome is computed by element-wise averaging). **As the optimal number of PACE-derived communities is 8 in F1000 and 9 in HCP (with a relatively small fifth community, C5, shown in Fig 3e) while Q-based methods primarily yield 3-5 communities, we selected a comparable PACE level, up to level 3, for our analyses.**

**Table 1** lists six Q-based methods adopted in this study (five weighted and one binarized). We conducted 100 runs for each of the six methods as well as PACE, and quantified pairwise similarity between two modular structures using NMI. We report summary statistics of these pairwise NMI values in **Table 2** (the total number of NMI values are 4950=100x99/2). As shown in this table, Q-based methods produced substantially variable modular structures across runs (and the number of communities across runs is also variable). By contrast, PACE produced identical results up to the third level (i.e., 8 communities) for HCP and F1000.

To visualize these modularity results, we show axial slices of representative modular structures, for the HCP dataset, generated using different methods (**Figure 4,** also see **Figure 3e** for rearranged connectome matrices based on PACE).

**Table 1** Summarizes the six Q-based methods, as implemented in the BCT toolbox, tested and compared in this study (Rubinov and Sporns, 2011, Schwarz and McGonigle, 2011, Betzel et al., 2016).

| Method | | Equation | | |
|---|---|---|---|---|
| Weighted version | Q-Comb-Sym | $Q = C^+Q^+ - C^-Q^-$ $\quad C^+ = C^-$ $Q^+ = f(W_{ij}^+)$ $\quad W_{ij}^+ = \begin{cases} W_{ij} & if\ W_{ij} > 0 \\ 0 & otherwise \end{cases}$ $Q^- = f(W_{ij}^-)$ $\quad W_{ij}^- = \begin{cases} -W_{ij} & if\ W_{ij} < 0 \\ 0 & otherwise \end{cases}$ | | |
| | Q-Comb-Asym | $Q = C^+Q^+ - C^-Q^-$ $\quad C^+ \neq C^-$ | | |
| | Q-Positive-only | $Q = f(W_{ij}^+)$ $\quad W_{ij}^+ = \begin{cases} W_{ij} & if\ W_{ij} > 0 \\ 0 & otherwise \end{cases}$ | | |
| | Q-Amplitude | $Q = f(|W_{ij}|)$ $\quad |W_{ij}| = \begin{cases} W_{ij} & if\ W_{ij} > 0 \\ -W_{ij} & otherwise \end{cases}$ | | |
| | Q-Negative-only | $Q = f(W_{ij}^-)$ $\quad W_{ij}^- = \begin{cases} -W_{ij} & if\ W_{ij} < 0 \\ 0 & otherwise \end{cases}$ | | |
| Binarizing | Thresholding | $Q = f(B_{ij})$ $\quad B_{ij} = \begin{cases} 1 & if\ W_{ij} > thres \\ 0 & otherwise \end{cases}$ | | |

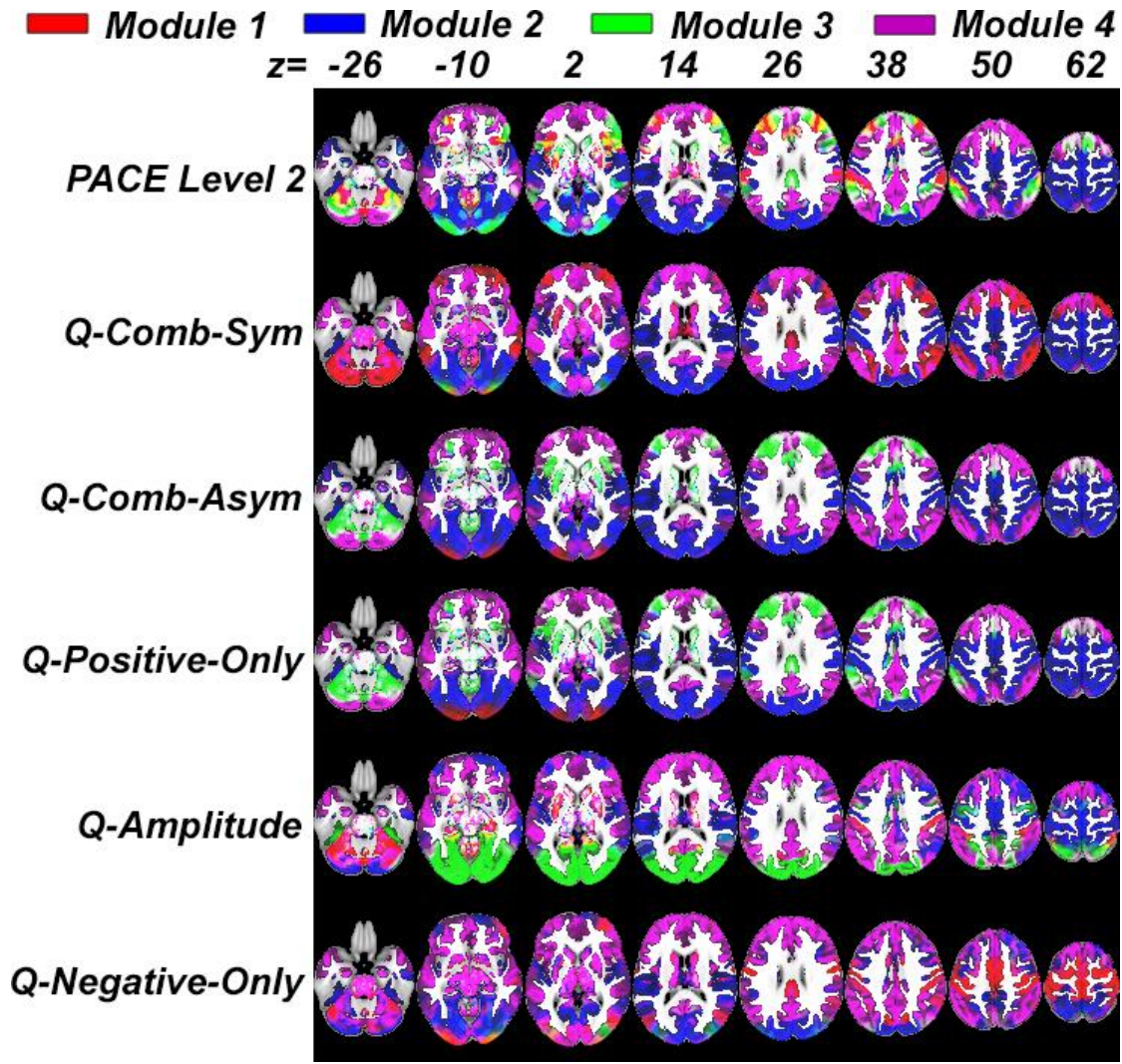

**Figure 4**. Representative modular structures generated using different methods for the HCP dataset. Regions coded in the same color (out of four: green, blue, red, and violet) form a distinct community or module. Note that unlike F1000, which uses structure parcellation to partition networks into non-overlapping communities, HCP utilizes an ICA-based parcellation, which allows components (modules) to overlap (Sporns and Betzel, 2016), resulting in regions with mixed colors (e.g. yellow).

**Table 2.** Mean and standard deviation of pair-wise Normalized Mutual Information (NMI) across 100 repeated runs within each method. The first three rows are from PACE and the rest from Q. For Q-based methods, the most reproducible methods are highlighted in bold (for F1000 it was the Q-Comb-Sym, and for HCP the Q-Positive-only).

| Method | F1000 | | HCP | |
|---|---|---|---|---|
| | NMI | Number of Modules (Number of runs) | NMI | Number of Modules (Number of runs) |
| PACE Level 1 | 1.0±0.0 | 2(100) | 1.0±0.0 | 2(100) |
| PACE Level 2 | 1.0±0.0 | 4(100) | 1.0±0.0 | 4(100) |
| PACE Level 3 | 1.0±0.0 | 8(100) | 1.0±0.0 | 8(100) |
| Q-Comb-Sym | **0.896±0.093** | 3(97),4(3) | 0.731±0.160 | 3(38), 4(62) |
| Q-Comb-Asym | 0.835±0.091 | 3(63),4(37) | 0.772±0.134 | 3(31),4(69) |
| Q-Positive-only | 0.844±0.103 | 3(1),4(99) | **0.834±0.079** | 3(41),4(59) |
| Q-Amplitude | 0.819±0.108 | 4(18),5(74),6(8) | 0.614±0.135 | 3(1),4(49),5(36),6(14) |
| Q-Negative-only | 0.617±0.158 | 3(66),4(34) | 0.460±0.129 | 3(3),4(61),5(36) |

As Q-based methods yielded variable results (with variable number of communities, see **Table 2**), for a fair comparison we randomly select a four-community modular structure to visualize each of the five Q-based methods. Visually, except for the Q-Amplitude and Q-negative-only, Q-based results shared **strong** similarities with results generated using 2$^{nd}$-level PACE (variability among Q-based methods notwithstanding). **Table 3** summarizes, for each Q-based method, the mean and standard deviation of NMI between the 100 runs and $2^{nd}$-level PACE-derived modularity.

**Table 3.** For each Q-based method, this table summarizes the pair-wise NMI's mean and standard deviation between any of the repeated 100 runs and the $2^{nd}$-level PACE-derived modularity.

| Method | F1000 | HCP |
|---|---|---|
| Q-Comb-Sym | 0.725±0.026 | 0.576±0.051 |
| Q-Comb-Asym | 0.705±0.043 | 0.603±0.047 |
| Q-Positive-only | 0.740±0.061 | 0.539±0.035 |
| Q-Amplitude | 0.606±0.064 | 0.235±0.027 |
| Q-Negative-only | 0.170±0.013 | 0.113±0.017 |

Last, to better visualize the effect of variable numbers of modules in Q-based methods, we randomly selected one 3-community and one 4-community Q-derived HCP modular structure and compared them (Figure 5), with the visualizations supporting the potential issue of reproducibility with Q (for comparison, the $1^{st}$ level 2-community and $2^{nd}$-level 4-community PACE HCP results are also shown).

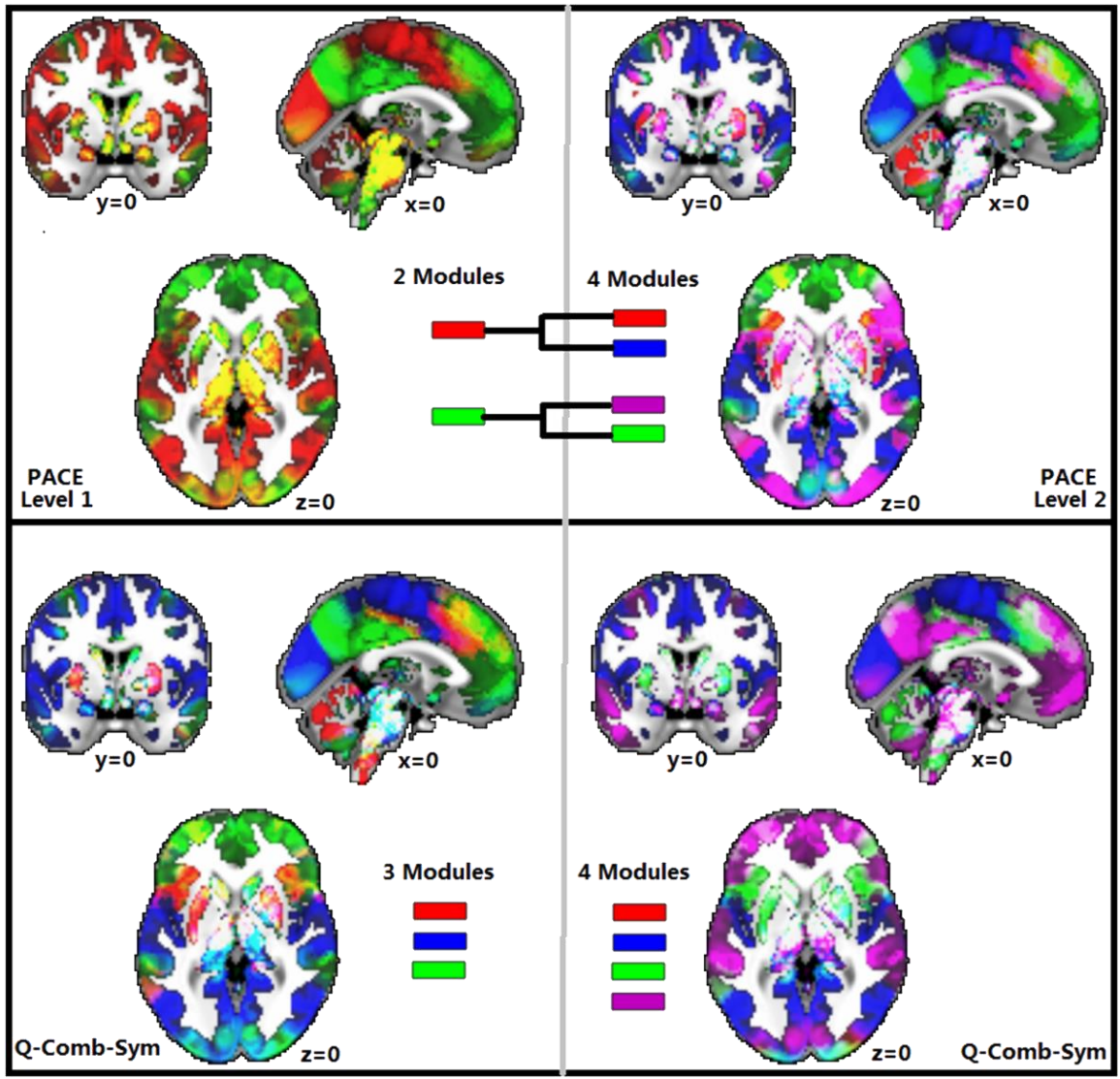

**Figure 5** Visualization of randomly selected 3-community and 4-community Q maximization-derived modular structures in HCP, demonstrating the suboptimal reproducibility with Q maximization. For comparison, the 1st level 2-community and 2nd-level 4-community PACE results are also shown.

## 3.6 Variability in the modular structure computed using Q-based thresholding-binarizing method

For the sixth Q-based modularity method, which applies an arbitrary non-negative threshold to the mean connectome followed by binarization (all edges below threshold set to zero, and above threshold to one), we again conducted 100 runs for each threshold (starting, as a fraction of the maximum value in the mean functional connectome, from 0 to 0.5 with increments of 0.02) using the un-weighted Louvain method routine implemented in the BCT toolbox. **Figure 6** plots the mean pairwise NMI ± SD as a function of the threshold, between each of the 100 runs and those generated using the Q-Comb-Sym or Q-Comb-Asym methods. Results again **demonstrated the substantial variability in Q as we vary the threshold**, especially in the case of HCP.

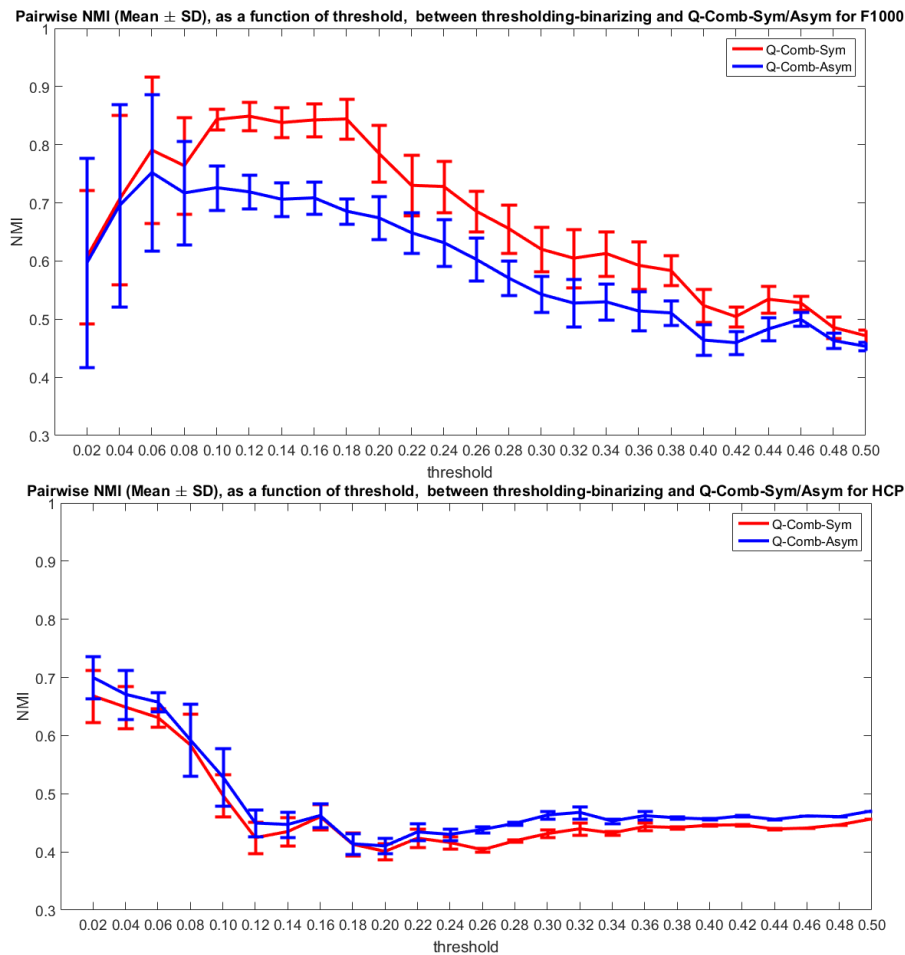

**Figure 6.** Mean and standard deviation of pair-wise similarity metric NMI, as a function of the threshold (x-axis, as a fraction of the maximal value in the mean group connectome),

between the modularity extracted using Q-based thresholding-binarizing and the weighted Q-Comb-Sym method or the Q-Comb-Asym method for F1000 (top) and HCP (bottom).

**3.7 Comparison between PACE modularity and Spectral Graph Cut**

Last, for completeness we also evaluated a network clustering algorithm derived from spectral graph theory (the normalized spectral cut or Ncut) (Ng et al., 2001). Since the Ncut algorithm only deals with positive edges and one needs to pre-specify the value of "k" (the number of clusters to be generated), we artificially set all negative edges to zero in the network and ran Ncut 100 times for k = 2, 4, and 8. Our results revealed that clustering derived from Ncut was also variable, as evidenced by the mean/standard deviation of pairwise NMI between any two of the 100 runs (Table 4). Note this should not come as a surprise, since the Ncut algorithm requires a random initialization step during the k-means step (i.e., even after k is determined, results are still dependent on how one initializes the center locations of the k clusters).

Further, we also computed the NMI between each of the 100 Ncut-derived modularity and that from PACE and reported the results in Table 5, which suggests substantial differences between the two.

**Table 4** The stability/variability of the Ncut algorithm, which was run 100 times for a k value of 2, 4, and 8 (each run is different due to the random initialization during the k-means step). We reported the pairwise NMI's mean and standard deviation between any two of the 100 runs for each k value.

|  | F1000 | | | HCP | | |
|---|---|---|---|---|---|---|
|  | 2 Modules | 4 Modules | 8 Modules | 2 Modules | 4 Modules | 8 Modules |
| Ncut | 0.9308± 0.0612 | 0.9374± 0.0814 | 0.6963± 0.0703 | 0.9469± 0.0520 | 0.6830± 0.1614 | 0.7184± 0.0735 |

**Table 5.** Comparing modularity derived from the Ncut algorithm and from PACE. Ncut was run 100 times with random initialization for a k value of 2, 4, and 8. We reported the pairwise NMI (mean and standard deviation) between any of the 100 runs and the corresponding PACE output (level 1 to 3, corresponding to 2, 4 and 8 modules).

|  | F1000 | | | HCP | | |
|---|---|---|---|---|---|---|
|  | 2 Modules | 4 Modules | 8 Modules | 2 Modules | 4 Modules | 8 Modules |
| NMI between Ncut and PACE | 0.4369± 0.0065 | 0.6635± 0.0322 | 0.6388± 0.0580 | 0.3745± 0.0133 | 0.5284± 0.0395 | 0.4529± 0.0199 |

## 3.8 Sex differences in resting-state networks using a PACE-based hierarchical permutation procedure

Because the HCP dataset has a better spatial resolution (2mm$^3$) and thus better suited for detecting modularity differences at a granular level (Van Essen et al., 2012, Van Essen et al., 2013), we demonstrate here that the stability of PACE makes it possible to pinpoint modularity differences between males and females in the HCP dataset, whilst minimizing potential confounding influences of age. As PACE uses a hierarchical permutation procedure to create trees, controlling for multiple comparisons is straightforward. Here, if two modular structures exhibit significant differences at each of the *m* most-local levels of modular hierarchy (each of them controlled at 0.05), collectively it would yield a combined false positive rate of 0.05 to the power of m. For the actual permutation procedure, we first computed the NMI between the two PACE-derived modular structures generated from the 367 males and the 453 females in the HCP dataset. Then, under the null hypothesis (no sex effect) we randomly shuffled subjects between male and female groups and recomputed the NMI between the permuted groups across all three levels of PACE-derived modularity. This shuffling procedure was repeated 10,000 times and the re-sampled NMI values were recorded.

By ranking our observed NMI among the re-sampled 10,000 NMI values, we detected significant sex differences in modularity starting at the first-level (*P* values: <1e-04, 1e-04, and 1.4e-03 for hierarchical level 1 to 3 respectively; a combined *P* value would thus be in the scale of $10^{-11}$). By contrast, a similar strategy to detect sex effect using any of the Q-based methods failed to identify significant differences in the two sex-specific modular structures. **Figure 7** visualizes the PACE-identified modular structure sex differences (highlighted using blue arrows and rectangles) in HCP.

**Figure 7** shows sex differences primarily in the bilateral temporal lobes, which was not detected using Q-based methods. These differences extended to the hippocampus and amygdala, which in females, were part of the green module, and in males formed part of the red module.

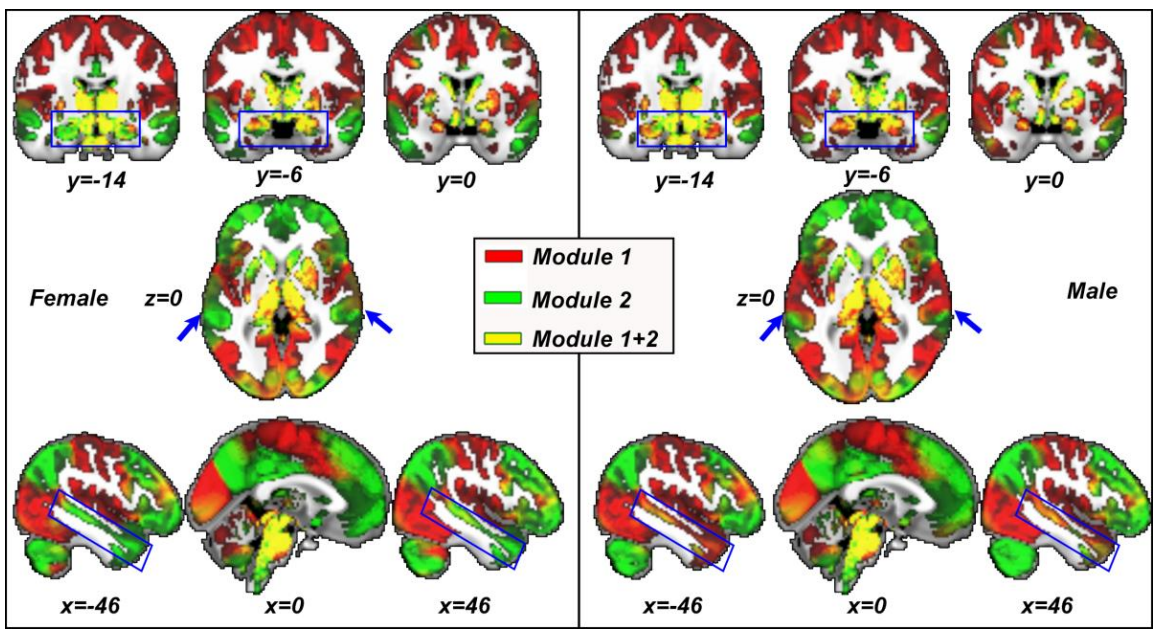

**Figure 7.** Visualization of PACE-identified sex-specific resting-state network modularity in females (left) and males (right) from the HCP dataset. Using permutation testing, sex-specific modularity differences are confirmed to be statistically significant throughout the entire PACE modular hierarchy starting at the first level. Here the results are visualized at first-level PACE, yielding two modules coded in red (module 1) and green (module 2). As HCP utilizes an ICA-based parcellation, modules thus overlap, in this case resulting in some regions colored in yellow.

Last, to validate these modularity findings we further replicated our hierarchical permutation procedure using a subset of F1000 in the age range of 20~30 (319 females at 23.25±2.26 years of age and 233 males at: 23.19±2.35), with results yielding not only visually highly similar PACE modularity (despite that F1000 and HCP are based on completely different brain parcellation techniques), but also similar sex differences (**Figure 8**; statistically significant at PACE level 1; p = 0.0378) in the limbic system (including the hippocampus) and the frontotemporal junction (including the pars opercularis as part of the inferior frontal gyrus), here primarily lateralized to the right hemisphere.

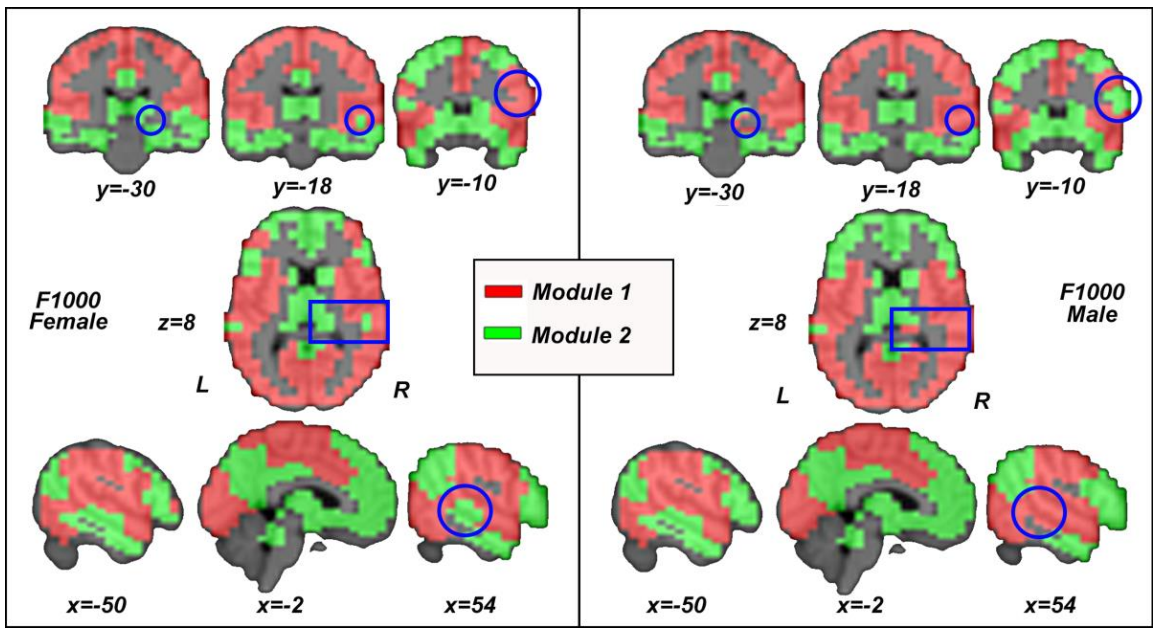

**Figure 8.** Visualization of PACE-identified resting-state network modularity and its sex differences between females (left) and males (right) from the F1000 dataset. The difference in spatial resolutions notwithstanding, note the visually highly similar PACE-

derived modularity in HCP (**Figure 7**) and F1000, even though these two datasets utilize completely different brain parcellation techniques.

## 4. Discussion

In this study, we proposed PACE, a novel way of understanding how anti-correlations help define modularity of the resting-state fMRI connectome. The benefit function to be optimized exploits the intuition that a higher probability of an edge being anti-correlated indicates a higher probability of it connecting regions in different modules. Importantly, PACE permits a symmetric equivalent dual form, such that it can be equally conceived as placing edges that are most consistently positive within modules. Thus, PACE is intrinsically symmetrized.

Conventional Q-maximization methods take a variable approach at negative edges. For example, many studies to date simply ignore anti-correlations by setting any values below a threshold (usually positive) to be zero (Sporns and Betzel, 2016), while others have proposed to down-weight negative edges in a somewhat heuristic fashion. The PACE method offers a novel and **theoretically advantageous** interpretation of left-tail fMRI networks, as traditionally, the left-tail network, i.e., those formed by negative edges alone, has been **at times** considered to be weak correlations that may "compromise" network attributes. For example, Schwarz and McGonigle (2011) argued that the left tail networks may not be biologically meaningful, despite noting that some connections were consistently observed in the negative-most tail networks, both with and without global signal removal. Schwarz and McGonigle (2011) thus recommended a "soft thresholding" approach be taken by replacing the hard thresholding or binarization operation with a continuous mapping of all correlation values to edge weights, suppressing rather than removing weaker connections and avoiding issues related to network fragmentation.

Rather than taking an approach that interprets the magnitude of correlations as the strength of connectivity, PACE determines the probability of a correlation being positive or negative. Interestingly, PACE can also be thought of as a different way of binarizing, with a "two-way" thresholding at zero. Although thresholding at a different value is possible, it would compromise the equivalence of the PACE dual forms. Indeed, one could theoretically generalize PACE by setting $P^{\alpha\pm}{}_{i,j}$ to compute the probability of $e_{i,j}$ being larger or smaller than an arbitrary threshold $\alpha$. However, in the case of a positive $\alpha$, the left tail is no longer strictly anti-correlations.

To validate PACE, we used full rather than partial correlations. We chose this method because recent literature has suggested that in general, partial correlation matrices need to be very sparse (Peng et al., 2009), and partial correlations have a tendency to reduce more connections than necessary. In dense networks such as fMRI-based brain networks, partial correlations have not been shown to be necessarily better than the Pearson correlation. Thus, partial correlations are often used in small networks that a) have small numbers of connections or b) have been forced to be sparse by introducing a penalty term during whole-brain network generation (Lee et al., 2011).

Following current practice in the literature, we compared PACE to Q maximization based modular structures in both the HCP and F1000 datasets using the default setting in the BCT toolbox (the Louvain method). **Notably, for Q-maximization we observed more variable modular structure, not only across different Q-based formulations (right tail, left tail, absolute value, and symmetric and asymmetric combined), but also across multiple runs within each formulation.**

A secondary analysis further applied PACE to the investigation of potential sex differences in the resting functional connectome. Note, while sex differences have been reported in the structural connectome of the human brain (e.g. Szalkai et al., 2015), few studies have examined sex differences in the functional connectome in healthy individuals, and no studies to our knowledge have examined sex differences in higher-level connectome properties such as network modularity. Previously, one large study (Biswal et al., 2010) examined the functional connectome of the F1000 dataset using three methods: seed-based connectivity, independent component analysis (ICA) and frequency domain analyses. Across the three analytic methods, they found consistent effects of sex, with evidence of greater connectivity in males than females in the temporal lobes, more so in the right hemisphere, and particularly when using ICA. Our results (Figure 8) are consistent with these reported findings.

Using the HCP dataset, our study also revealed higher-level sex-specific connectome modularity differences in the temporal lobes, including the middle temporal gyrus, amygdala and hippocampus. The amygdala (Cahill, 2010) and hippocampus (Addis et al., 2004) are important for emotional and autobiographic memory, while previous studies have reported sex differences in their activities in this context (Seidlitz and Diener, 1998, Davis, 1999, St. Jacques et al., 2011, Young et al., 2013). In line with these findings that likely reflect differential, sex-specific cognitive strategies for recalling memories related to self, we found that in female, the amygdala and hippocampus are within the module that also contains the default mode network, whereas in males they belong to the module largely consisting of the visual and somatomotor networks. It is also consistent with Damoiseaux et al (2016), which found that females had greater connectivity between the hippocampus and medial PFC than males, and Kogler et al (2016) which found that females had greater connectivity between the left amygdala and left middle temporal gyrus than males. These medial prefrontal and lateral temporal regions form part of the brains default mode network (Fox et al., 2005). Thus, these recent and preliminary findings may reflect stronger coupling within the default mode network and between the amygdala and the default mode network in females than in males, supporting previous reports of greater regional homogeneity in the right hippocampus and amygdala in females than males (Lopez-Larson et al., 2011).

## 5. Limitations and future directions

**First, we note that the proposed PACE framework is based on the estimation of edge positivity/negativity frequency, which encodes details of functional co-activation/anti-activation. Unlike Q-based methods that encode such details using correlation magnitudes, PACE procedure discards edge weights, which may be part of the reason why it yields more**

stable results by discarding otherwise useful details and reducing accuracy. However, a counter argument can also be made in that a majority of noisy features tend to be close to zero with arbitrary signs; thus Q-based methods that employ thresholding can simply remove these noisy features and probably yield more stable results (although this is not supported by our thresholding-binarization experiments in section 3.6).

Further, the idea that correlation magnitudes always encode meaningful details is also not universally accepted in the imaging community, as evidenced by studies that instead adopted a thresholding-binarization approach. For example, in (van den Heuvel et al., 2017), the authors extensively tested a proportional threshold approach that "includes the selection of the strongest PT% of connections in each individual network, setting all (in the binary case) surviving connections to 1 and other connections to 0" in order to "remove spurious connections and to obtain sparsely connected matrices, a prerequisite for the computation of many graph theoretical metrics"

Second, with the PACE benefit function cast as a difference between inter-modular versus intra-modular mean edge negativity, the optimization problem is NP-hard and thus the global solution is not computable in realistic terms. Thus, we instead used a top-down hierarchical bifurcating solver that was previously extensively tested in PLACE. Despite this limitation, we a) outlined a theoretical connection between PACE and the Ising model, demonstrating that the PACE algorithm is a maximum likelihood **estimation algorithm, b) showed that PACE results were robust and insensitive to multiple runs while recovering known resting-state networks**, **and c) showed that PACE-derived number of communities is not restricted to powers of 2, due to a permutation procedure that constructs the null distribution of the observed data allowing us to determine, at each branch, if further bifurcation is statistically meaningful**.

Although the novel PACE-based symmetrized functional modularity is shown to be a powerful and mathematically elegant approach to understanding anti-correlations in fMRI connectomes, it cannot be computed without robust estimates of edge negativity/positivity frequencies**, and thus there may be instances where Q yields more biologically meaningful results.** Here we tested PACE using large-N cohorts of HCP and F1000 functional connectomes. Although theoretically feasible, individual-level PACE would require multiple runs for each individual or alternatively a completely different mathematical formulation (e.g., non-correlation based, see below) for estimating these frequencies.

**Along this line, we note that recently several more sophisticated approaches for rigorous null modeling of correlation matrices and for multilayer multiscale Q maximization have been proposed (Betzel et al., 2015, MacMahon and Garlaschelli, 2015, Bazzi et al., 2016, Betzel et al., 2016). For example, in MacMahon and Garlaschelli (2015) the authors used random matrix theory to identify non-random properties of empirical correlation matrices, leading to the decomposition of a correlation matrix into a "structured" and a "random" component. While beyond the scope of this**

**study, we note that 1) such a decomposition requires strong assumptions that have been criticized and may not hold for the human brain, 2) PACE extracts modularity given *some estimation* of functional co-activation/anti-activation via edge positivity/negativity, which can be based on time series correlation (an approach we adopted here due to its conventional popularity), based on more advanced null modeling as in this cited study, or based on other information-theoretical approaches that we are currently exploring and are completely non correlation-based. Thus, individual subject-level PACE becomes possible.**

Last, a new multi-scale modularity maximization approach has been recently investigated that seeks to generalize the Q modularity metric, and is thus likely to outperform the Q methods we studied here. However, in contrast to the simplicity of the PACE model (that does not require any parameter tuning) and in addition to the several caveats and nuances of the existing Q methods, this multi-scale approach introduces additional resolution parameter ($\gamma$) that needs to be further tuned (a range from $10^{-2.0}$ to $10^0$ was studied in Betzel et al. (2015))

Notwithstanding the several limitations of correlation-based PACE noted above, we demonstrated that testing specific effects (e.g., sex) can be achieved with careful permutation testing while controlling for other variables (such as age), as in our secondary analyses showing significant sex effects in the temporal lobes. Lastly, a recent study has utilized the rich dataset provided by the HCP to develop a new multimodal method for parcellating the human cerebral cortex into 180 areas per hemisphere (Glasser et al., 2016). This semi-automated method incorporates machine-learning classification to detect cortical areas. It would be interesting to apply PACE to this new parcellation once the classifier becomes publicly available.

*6. Conclusions*

**This methodological report outlines a novel PACE framework that complements the existing Q-based methods of defining modularity for brain networks in which negative edges naturally occur.** When applied to the HCP and the F1000 datasets, **we showed that PACE yielded stable reproducible results that are consistent with those derived from existing methods**, providing evidence for convergent validity. Furthermore, given the high reliability of this new method, we have been able to demonstrate sex differences in resting state connectivity that are not detected with traditional methods.